\documentclass{aa} \usepackage{txfonts}
\usepackage{graphicx}

\begin{document}

\title{The hot subdwarf B + white dwarf binary KPD~1930+2752}

\subtitle{a Supernova Type Ia progenitor candidate
\thanks{Some of the data presented here were obtained at the W.M. Keck Observatory, which is operated as a scientific partnership among the California Institute of Technology, the University of California and the National Aeronautics and Space Administration. The Observatory was made possible by the generous financial support of the W.M. Keck Foundation.} \fnmsep 
\thanks{Based on observations collected at the Centro Astron\'omico Hispano Alem\'an (CAHA) at Calar Alto, operated jointly by the Max-Planck Institut für Astronomie and the Instituto de Astrof\'isica de Andaluc\'ia (CSIC).} \fnmsep
\thanks{Based on observations at the Paranal Observatory of the European Southern Observatory for programme 167.D-0407.}
\thanks{R. N. gratefully acknowledges support by a PPARC Advanced Fellowship.}
}

\author{S. Geier \inst{1}
   \and S. Nesslinger \inst{1}  
   \and U. Heber \inst{1}  
   \and N. Przybilla \inst{1} \and R. Napiwotzki \inst{2} \and R.-P. Kudritzki \inst{3}}

\offprints{S.\,Geier,\\ \email{geier@sternwarte.uni-erlangen.de}}

\institute{Dr.--Remeis--Sternwarte, Institute for Astronomy, University Erlangen-Nuremberg, Sternwartstr. 7, 96049 Bamberg, Germany 
   \and Centre of Astrophysics Research, University of Hertfordshire, College Lane, Hatfield AL10 9AB, UK
   \and Institute for Astronomy, University of Hawaii, 2680 Woodlawn Drive, Honolulu, HI 96822, USA}

\date{Received \ Accepted}

\abstract{
{\it Context.}
The nature of the progenitors of type Ia supernovae is still under controversial
debate. KPD~1930+2752 is one of the best SN Ia progenitor candidates known
today. The object is a double degenerate system consisting of a
subluminous B star (sdB) and a massive white dwarf (WD). Maxted et al. (\cite{maxted})
conclude that the system mass exceeds the Chandrasekhar mass. This
conclusion, however, rests on the assumption that the sdB mass is
$0.5\,M_{\odot}$. However, recent binary population synthesis calculations
suggest that the mass of an sdB star may range from $0.3\,M_{\odot}$ to more
than $0.7\,M_{\odot}$.\\ 
{\it Aims.}
It is therefore important to measure the mass of the sdB star simultaneously with that
of the white dwarf.
Since the rotation of the sdB star is tidally locked to the orbit the 
inclination of the system can 
be constrained if the sdB radius and the projected rotational velocity 
can be measured with high precision. An analysis of the ellipsoidal variations in the 
light curve allows to tighten the constraints derived from spectroscopy.\\
{\it Methods.}
We derive the mass-radius relation for the sdB star from 
a quantitative spectral analysis of 150 low-resolution spectra  
obtained with the Calar Alto 2.2 m telescope using metal-rich, line blanketed 
LTE model atmospheres with and without NLTE line formation.
The projected rotational 
velocity is determined for the first time 
from 200 high-resolution spectra obtained with the Keck I 10 m, and 
the ESO-VLT 8.2 m telescopes. In addition a reanalysis of the published light curve 
is performed.\\
{\it Results.}
The atmospheric and orbital parameters are measured with unprecedented accuracy. In particular the projected rotational velocity \(v_{\rm rot}\sin{i} = 92.3 \pm 1.5~{\rm km\,s^{-1}}\) is determined.
Assuming the companion to be a white dwarf, the mass of the sdB is limited 
between \(0.45~M_{\odot}\) and \(0.64~M_{\odot}\) and the corresponding total 
mass of the system ranges from \(1.33~M_{\odot}\) to \(2.04~M_{\odot}\). This
constrains the inclination to $i>68\degr$. The photometric analysis allows to constrain the parameters even more.
A neutron star companion can be ruled out and the mass of the sdB is limited between
\(0.45~M_{\odot}\) and \(0.52~M_{\odot}\). The total mass of the system ranges from
\(1.36~M_{\odot}\) to \(1.48~M_{\odot}\) and hence is likely to exceed the Chandrasekhar mass. The inclination angle is
$80\degr$ and the light curve shows weak and shallow signs of eclipses. A high precision light curve is needed in order to accurately measure these eclipses.
So KPD~1930+2752 qualifies as an excellent double degenerate supernova Ia progenitor candidate. 

}

\maketitle

\section{Introduction}

Supernovae of type Ia (SN~Ia) play an important role in the study of cosmic 
evolution. They are regarded as the best standard candles for the determination of the cosmological parameters $H_{\rm 0}$, $\Omega$ and $\Lambda$ (e.g. Riess et al. \cite{riess}; Leibundgut \cite{leibundgut}; Perlmutter et al. \cite{perlmutter}) together with the WMAP probe for anisotropies in the cosmic background radiation (e. g. Bennett et al. \cite{bennett}). Although some of the most important discoveries in modern cosmology, in particular the accelerated expansion of the universe, were initially derived from distance measurements of SN~Ia with high redshifts, the nature of their progenitors is still under debate (Livio \cite{livio}). The progenitor population is a crucial information required to back the assumption that distant SN~Ia can 
be used as standard candles as it was done with the ones in the local universe.
There is general consensus that only the thermonuclear explosion of a 
white dwarf (WD) is compatible with the observed features of SN~Ia. 
For this a white dwarf has to accrete mass from a close companion to reach 
the Chandrasekhar limit of $1.4 \,M_{\rm \odot}$ (Hamada \& Salpeter \cite{hamada}). Two main scenarios of 
mass transfer are currently under discussion. 
In the so-called single degenerate (SD) scenario (Whelan \& Iben \cite{whelan}),
 the mass donating component is a red giant/subgiant, which fills its Roche 
 lobe and is continually transfering mass to the white dwarf. According to 
 the so-called double degenerate (DD) scenario (Iben \& Tutukov \cite{iben}) 
 the mass donating companion is a white dwarf, which eventually merges 
 with the primary due to orbital shrinkage caused by gravitational wave 
 radiation.
 
A progenitor candidate for the DD 
scenario must have a total mass near or above the Chandrasekhar limit and 
has to merge in less than a Hubble time. 
Systematic radial velocity (RV) searches for DDs have been undertaken 
(e.g. Napiwotzki \cite{napiwotzki2} and references therein). 
The largest of these projects was the 
ESO SN Ia Progenitor Survey (SPY). More than $1\,000$ WDs were checked for 
RV-variations (e. g. Napiwotzki et al. \cite{napiwotzki2}). SPY detected 
$\sim 100$ new DDs (only $18$ were known before). One of them may fulfill 
the criteria for progenitor candidates, even though the error margins are 
quite high (Napiwotzki et al. \cite{napiwotzki3}; Karl \cite{karl}).

KPD~1930+2752 was identified as a subdwarf B (sdB) star in the Kitt Peak -- 
Downes survey (Downes \cite{downes}). The parameters which were derived from 
spectroscopy by model atmosphere fits (Saffer \& Liebert \cite{saffer}) are 
consistent with the theoretical instability strip predicted by 
Charpinet et al. (\cite{charpinet}). After the first pulsating sdB stars 
with short periods (EC 14026 stars) were 
discovered (Kilkenny et al. \cite{kilkenny}; Koen et al. \cite{koen}; 
Stobie et al. \cite{stobie}), Bill\`{e}res et al. (\cite{billeres}) 
initiated a survey to search for such objects in the northern hemisphere. 
KPD~1930+2752 was selected from the list of Saffer \& Liebert (\cite{saffer}) 
for their fast photometry program. They detected multiperiodic variations with 
short periods and low amplitudes. In addition to 44 $p$-mode pulsations they 
found a strong variation at a much longer period of about \(4100 \, {\rm s}\). 
This variation could be identified as an ellipsoidal deformation of the sdB 
star most likely caused by a massive companion. Bill\`{e}res et al. (2000) 
predicted the period of the binary to be two times the period of this 
brightness variation (\(P=8217.8 \, {\rm s} = 0.095111 \, {\rm d}\)).

This was proven by Maxted et al. (\cite{maxted}), who measured a radial 
velocity curve of KPD~1930+2752 which matched the proper period. The radial 
velocity amplitude \(K=349.3 \pm 2.7 \, {\rm km\,s^{-1}} \) combined with an 
assumption of the canonical mass for sdB 
stars \(M_{ \rm sdB}=0.5 \, M_{\odot}\) led to a lower limit for the mass of 
the system derived from the mass function. 
This lower limit \(M \geq 1.47\, M_{\odot} \) exceeded the Chandrasekhar 
mass of \(1.4 \, M_{\odot}\) (Hamada \& Salpeter \cite{hamada}). Because there was no sign of a companion in 
the spectra, it was concluded that the unseen object must be a white dwarf. 
The system is considered to become double degenerate when the subdwarf 
eventually evolves to a white dwarf. Orbital shrinkage caused by 
gravitational wave radiation will lead to a merger of the binary in about 
$200$ Myr which is much less than a Hubble time. 
Combining all the evidence, Maxted et al. concluded that KPD~1930+2752 could 
be a good candidate for the progenitor of a Type Ia supernova.

From the theoretical point of view Ergma et al. (\cite{ergma}) questioned the 
double degenerate scenario in the case of KPD~1930+2752. Their simulations 
based on the derived parameters of Maxted et al. (\cite{maxted}) suggested 
that the merger will lead to a single massive ONeMg white dwarf rather than 
a supernova explosion. The merger of two white dwarfs is a rather complicated 
process compared to slow accretion of material onto a heavy white dwarf. Detailed merger
simulations in 3D are not yet avaible, but urgently needed to learn more about this
possible SN~Ia progenitor channel. 

The subluminous B stars form the hot end of the horizontal branch, the so
called Extreme Horizontal Branch (EHB, Heber \cite{heber1}). The EHB models 
imply
that they are core helium burning objects with tiny hydrogen dominated
envelopes, too thin to sustain hydrogen burning. The core
mass is fixed by the onset of the core helium flash at the tip of the red giant
branch and depends only slightly on metallicity and helium abundance. Hence the 
canonical core mass is resticted to a very narrow range of $0.46$ to $0.5\,M_{\odot}$. 
Unlike normal horizontal branch
stars, EHB stars do not ascend the asymptotic giant branch (AGB) after core 
helium 
burning has terminated, but evolve to higher temperatures until the white dwarf 
cooling sequence is reached.

Binary evolution scenarios have been proposed (e.g. Han et al., \cite{han1}, 
\cite{han2})
to explain the large fraction of
sdB stars in binaries. These scenarios involve strong mass-transfer either 
through a common-envelope ejection, or through stable Roche lobe overflow.
A merger of two helium core white dwarfs is another viable option.

The binary population synthesis models of Han et al. (\cite{han1}) predict a mass
distribution for sdB stars that peaks near the canonical mass, 
but is much wider, ranging from $0.3\,M_{\odot}$ to more 
than $0.7\,M_{\odot}$. The lower limit is based on the fact that no core helium 
burning 
can be sustained for stars with masses less than $0.3\,M_{\odot}$.
The highest masses result from mergers.
Hence it might be premature to assume that the mass of an sdB star
is half a solar mass, as has been done in previous studies.
Therefore we drop this assumption and treat the mass of the sdB component as a
free parameter.        

Another drawback of all previous investigations was the lack of information 
on the inclination angle. We aim at deriving constraints on the inclination 
for the first time by means of accurate measurements of the projected 
rotational 
velocity and the surface gravity from appropriate high-quality spectra using 
sophisticated model atmospheres. Because the rotation of the sdB star is 
tidally locked to the orbit, we can derive \(\sin{i}\) as a function of the 
sdB mass from these two quantities. This approach combined with a reanalysis of the published light curve allows to constrain the 
masses of both components and, therefore, to check
whether or not KPD~1930+2752 qualifies as a SN~Ia progenitor candidate through 
a double degenerate merger.   

Section \ref{sec:obs} describes the newly obtained spectra, from which an
improved radial velocity curve has been derived (Sect.~\ref{sec:rvc}).
For the quantitative spectral analysis, existing grids of metal line blanketed
LTE models have been used and a new (hybrid) approach is applied that 
allows to treat both departures from LTE as well as metal line blanketing
(Sect.~\ref{sec:atm}). Combining these results, described in 
Sect.~\ref{sec:par}, with the projected rotational velocity
(Sect.~\ref{sec:rot}) we constrain masses and inclination of the 
KPD~1930+2752 system (Sect. \ref{sec:mi}). Within the derived parameter space synthetic 
light curves were modelled and fitted to the published light curve (Sect.~\ref{sec:phot})
in order to constrain the parameters even further.
Our conclusions are given in the final 
section.   

\begin{figure}[t!]
	\centering
	\resizebox{\hsize}{!}{\includegraphics{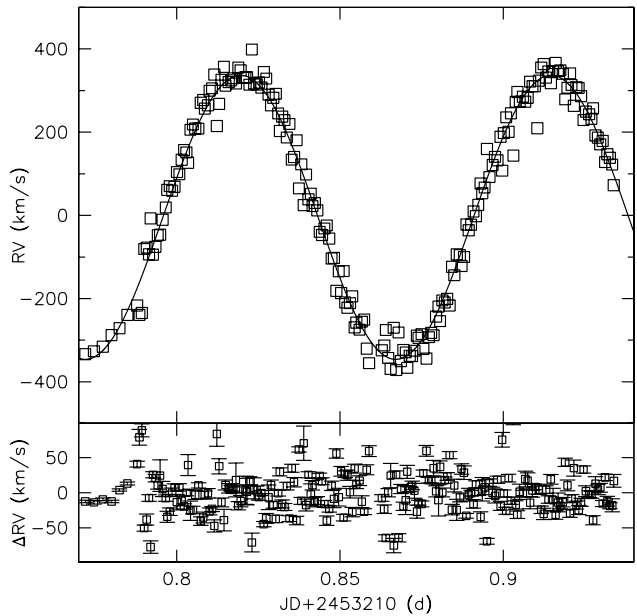}}
	\caption{Radial velocity curve. The data points are from the new HIRES spectra, the solid line is the fitted sine curve to these points. The residuals in the lower part show no signs of eccentricity.}
	\label{RV1}
\end{figure}

\section{Observations and Data Reduction \label{sec:obs}}

With the \(10 \, {\rm m}\) Keck I Telescope at the Mauna Kea Observatory two 
hundred high-resolution spectra were obtained by N. Przybilla in half a night 
in July 2004, using the High Resolution Echelle Spectrometer (HIRES; 
Vogt et al. \cite{vogt}). The spectra covered a wavelength range of 
\(4\,200 \, {\rm \AA} - 6\,600 \, {\rm \AA} \) with a few small gaps at a 
resolution of \(0.1 \, {\rm \AA} \) and exposure times of \(20 \, {\rm s}\) 
each. The data were reduced using the ESO-MIDAS package. Bias and flatfield 
corrections were applied. The echelle orders were extracted and sky background
was subtracted. The wavelength calibration was done separately for every single order.
All spectra were corrected to the heliocentric frame of reference. To perform model 
atmosphere fits all spectra were radial velocity corrected to rest wavelength and 
a coadded spectrum was generated separately for every echelle order.

Two spectra were taken with the ESO Very Large Telescope UT2 (Kueyen) equipped 
with the UV-Visual Echelle Spectrograph (UVES) in spring 
2001 (Dekker et al. \cite{dekker}). The spectra cover a wavelength range of 
$3300 - 6650 \, {\rm \AA}$ with two small gaps at a resolution of 
$0.1 \,{\rm \AA}$. The exposure time was $300 \, {\rm s}$ each. The spectra 
were reduced with a pipeline developed by C. Karl using the ESO-MIDAS package 
and based on the implemented UVES reduction pipeline (Karl \cite{karl}). 
Additional observations were obtained with the \(2.2 \, {\rm m}\) Telescope at 
the Calar Alto Observatory in July 2004. The Calar Alto Faint Object 
Spectrograph (CAFOS) was used to obtain 150 spectra covering a wavelength 
range of \(3\,600 \, {\rm \AA} - 6\,200 \, {\rm \AA} \) with 
\(5 \, {\rm \AA} \) resolution and an exposure time of \(180 \, {\rm s}\) 
each. The data were reduced and coadded in analogy to the HIRES spectra.

\begin{figure*}[t!]
	\resizebox{\hsize}{!}{\includegraphics{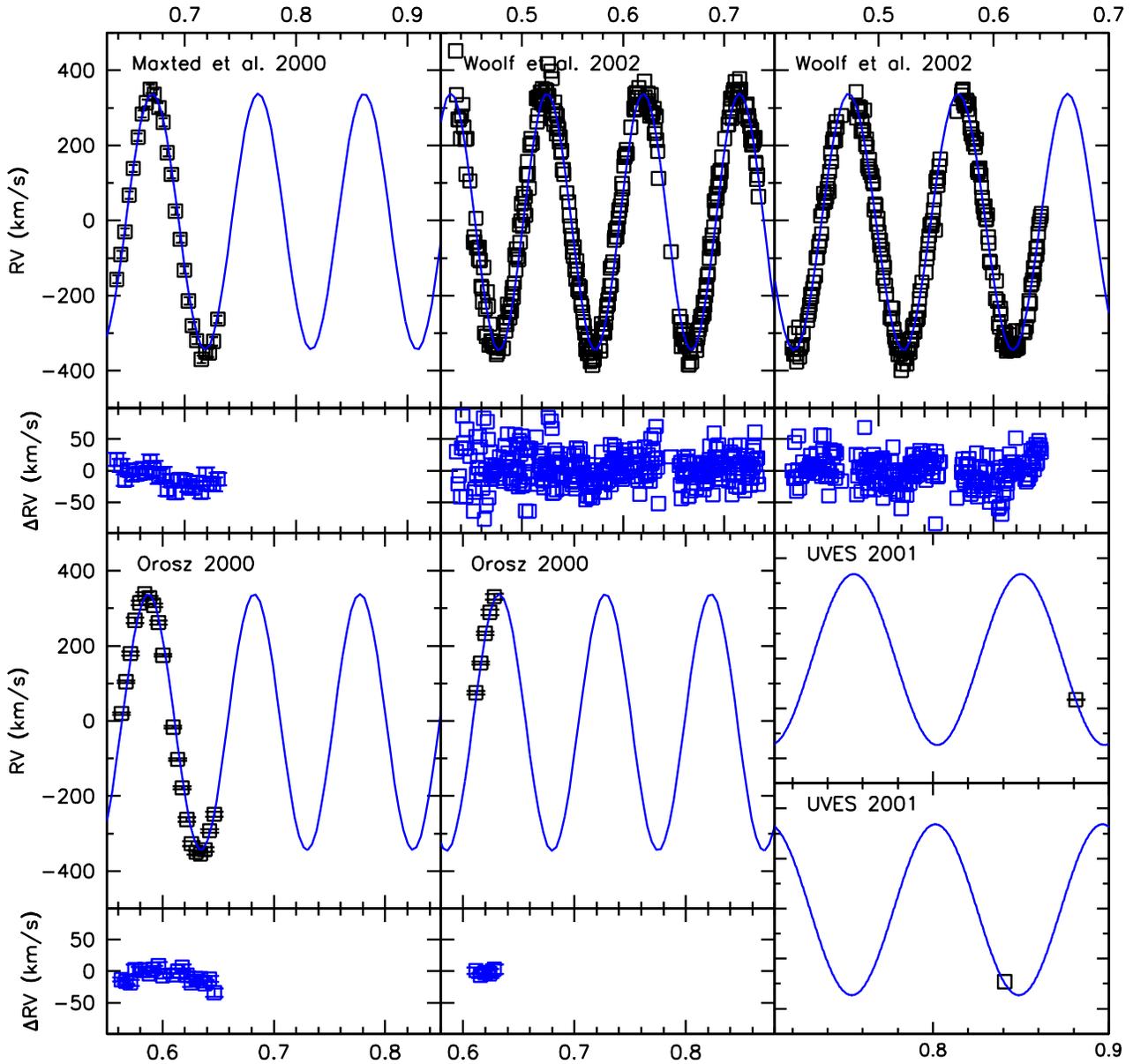}}
	\caption{Radial velocity curves for the years 2000 and 2001. Times are
	given in days. The corresponding HJD has to be added. 
	Data are from Maxted et al. (\cite{maxted}: HJD = 2451651); 
	Woolf et al. (\cite{woolf}: HJD = 2451716/2451717); Orosz (priv comm.: 
	HJD = 2451769/2451770), and from UVES spectra (HJD = 2452033/2452065). 
	The solid line is the fitted sine curve to all data points 
	over a timespan of four years. 
	The residuals in the lower parts show no signs of eccentricity. 
	The error bars in the curve of Woolf et al. have been skipped, 
	because they are very large due to the low S/N ratio of the spectra.}
	\label{RV2}
\end{figure*}

\section{Radial Velocities and Spectroscopic Orbit \label{sec:rvc}}

Because the S/N ratio of the individual HIRES spectra was very low, only the 
H$\alpha$ and H$\beta$ lines could be used for determining the radial velocity 
by \(\chi^{2}\) cross correlation with a model spectrum at rest wavelength. To 
improve the accuracy, the resulting radial velocity curve was combined with 
all available radial velocity data of KPD~1930+2752 
(Maxted et al. \cite{maxted}; Woolf et al. \cite{woolf}; 
Orosz 2000 priv. comm.) covering a time\-span of four years. A sine curve was 
fitted to these 2\,900 data points using a \(\chi^{2}\) minimizing method and 
the power spectrum was generated (FITRV and FITPOW routines by H. Drechsel; 
Napiwotzki et al. \cite{napiwotzki5}). The sine curve fit is excellent for all 
different datasets and no period change could be detected over the whole 
timebase (Figs. \ref{RV1} \& \ref{RV2}). The orbital parameters were measured 
with unprecedented accuracy and are consistent with prior measurements by 
Maxted et al. (\cite{maxted}) and Woolf et al. (\cite{woolf}): 
\(\gamma({\rm H}\alpha)=5 \pm 1 {\rm \,km\,s^{-1}}\), 
\(K=341 \pm 1 \,{\rm km\,s^{-1}}\), 
\(P=0.0950933 \pm 0.0000015~ {\rm d}\). The residuals do not show any signs of 
eccentricity. 

No radial velocity variations due to pulsations could be found
either. 
This result is consistent with the non-detection of RV variations by 
Woolf et al. (\cite{woolf}). The high orbital RV-variation of the system 
enforces short exposure times in order to prevent orbital smearing effects. 
Considering the fact that our spectra were obtained with the largest optical 
telescope in the world, it can be doubted that such a detection would be 
feasible for this object with present-day instrumentation.

\section{Atmospheric models and synthetic spectra \label{sec:atm}}

Up to now spectra of sdB stars were analysed either from 
metal line blanketed LTE model atmospheres or from NLTE model
atmospheres neglecting metal line blanketing altogether. As pointed out by 
Heber et al. (\cite{heber2}) and Heber \& Edelmann (\cite{heber4}) systematic 
differences 
between these two approaches are present. 
Most importantly the gravity 
scale differs by about 0.06 dex. 
O'Toole \& Heber (\cite{otoole}) found very high abundances of heavy elements in 
pulsating sdB stars of similar temperature as KPD~1930+2752 and showed that 
model atmospheres with supersolar metallicity (10 $\times$ solar) reproduced 
the optical spectra much better than solar metallicity models did.
Hence in the analysis of the spectra of KPD~1930+2752 we used their model 
grids for both solar and 10 times solar metallicity.

As the gravity is of utmost importance 
for our analysis we also calculated new grids of models and synthetic  
spectra to account for NLTE effects and metal line blanketing 
simultaneously. Since the temperature/density structure of an sdB 
atmosphere is only slightly affected by NLTE effects (if at all), 
the LTE approximation is valid to this end. 
NLTE effects may become more important for the 
line formation of Balmer and helium lines. 
Therefore, we chose a ``hybrid'' approach by calculating 
the temperature/density stratification from metal line blanketed LTE 
model atmospheres and then performing line formation calculations for hydrogen 
and helium allowing for 
departures from LTE (Przybilla et al. \cite{przybilla3}). We used both of the Kurucz codes, ATLAS9 and ATLAS12,
to calculate metal line blanketed model atmospheres (Kurucz~\cite{kurucz1}, 
\cite{kurucz2}).
The coupled statistical equilibrium and radiative transfer equations were 
solved and spectrum synthesis with refined line-broadening theories was 
performed using {\sc Detail} and {\sc Surface} (Giddings \cite{giddings}; 
Butler \& Giddings \cite{butler}).
 Both codes have undergone major revisions and improvements over the past 
 few years. State-of-the-art NLTE model atoms for hydrogen and helium are 
 utilised (Przybilla \& Butler \cite{przybilla1}; Przybilla \cite{przybilla2}).

The Stark broadening of hydrogen lines is important for the determination 
of temperature and gravity. Up to now all spectral analyses of sdB stars 
used synthetic Balmer line spectra based on the unified theory 
of Vidal, Cooper \& Smith (\cite{vidal}, VCS) with improvements of 
Lemke (\cite{lemke}). Since new 
broadening tables for Balmer lines have become available (Stehl\'e \& 
Hutcheon \cite{stehle}, SH), we also calculated synthetic spectra from these tables 
to investigate 
systematic effects on the synthetic spectra.

\begin{figure}[t!]
	\resizebox{\hsize}{!}{\includegraphics{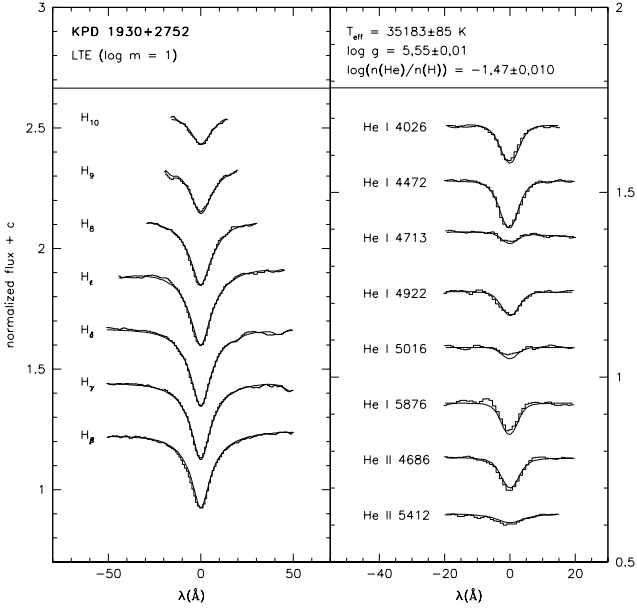}}
	\caption{Best fit with an LTE model of ten times solar metallicity 
	(model B of Table~\ref{tab:par}).}
	\label{LTEm1}
\end{figure}

\section{Atmospheric Parameter Determination \label{sec:par}}

The CAFOS spectra were preferred for the parameter determination because they 
are of excellent S/N and the run of the continuum is smooth. The continuum of 
the Keck HIRES spectrum suffers from a non-optimal rectification of the 
continuum, which rendered the analysis of the Balmer line wings difficult or 
unreliable. Therefore we restrained ourselves from determining atmospheric 
parameters from the Keck HIRES spectra.

The observed spectra were analysed with FITPROF by means of a 
\(\chi^{2} \) fit (Napiwotzki \cite{napiwotzki1}). Rotational broadening was 
accounted for by choosing an appropriate value for the projected rotational 
velocity (see Section 6).

Motivated by the discovery of strong $\log g$ variations in the pulsating 
subdwarf star PG 1325+101 (Telting \& \O stensen \cite{telting}), this 
possibility was also taken into account. To search for changes of the 
stellar parameters due to pulsations or due to the ellipsoidal deformation of 
the subdwarf star, three to ten CAFOS spectra were coadded. The atmospheric 
parameters for every phase were determined. No changes could be detected.
Then we combined all CAFOS spectra to form a very high S/N spectrum and performed a fit. The formal 
statistical errors were small, i.e. $\Delta T_{\rm eff} \approx 100 \,{\rm K}, 
\Delta \log\,g \approx 0.005\, {\rm dex}, \Delta \log\,\frac{n(\rm He)}{n(\rm H)} \approx 0.005\, {\rm dex}$. 
However, systematic errors due to inaccuracies in the model atmospheres and 
synthetic spectra may be much larger. 
To investigate this we used the different grids of models described in 
Section~\ref{sec:atm} to determine the atmospheric parameters.

\begin{figure}[t!]
	\resizebox{\hsize}{!}{\includegraphics{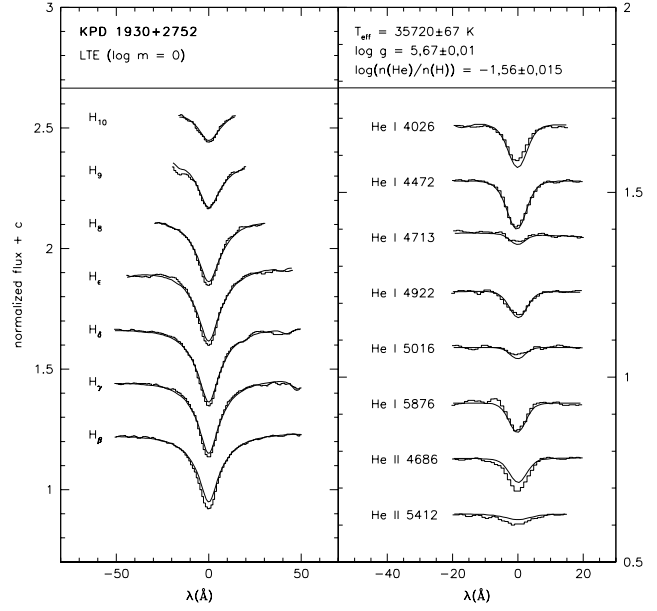}}
	\caption{Fit with an LTE model grid of solar metallicity (model A 
	of Table~\ref{tab:par}).}
	\label{LTEm0}
\end{figure}

\begin{table}[h!]
\caption{Atmospheric Parameters of KPD~1930+2752 derived from model atmosphere 
fits. A12=Kurucz ATLAS12 code, VCS=Vidal, Cooper \& Smith (\cite{vidal}), 
SH=Stehl\'e \& Hutcheon (\cite{stehle}), lf=line formation.} 
\label{tab:par}
\begin{center}
\begin{tabular}{cllccc}
	\hline
	& model & $[m/H]$ & $T_{\rm eff}$ & $\log g$ & $\log \frac{n({\rm He})}{n({\rm H})}$\\
	\hline 
	A & LTE VCS & 0 & 35\,720 {\rm K}    & 5.67  &   -1.56\\
	B & LTE VCS & +1 & 35\,183 {\rm K}    & 5.55  &   -1.47\\
	C & NLTE (lf) VCS  	& +1 & 35\,353 {\rm K}    & 5.61  &   -1.50\\
	D & NLTE (lf) SH  & +1 & 35\,212 {\rm K}    & 5.67  &   -1.51\\
	E & NLTE (lf) SH A12  & +1 & 35\,712 {\rm K}    & 5.67  &   -1.58\\
	\hline \hline
	& adopted & & 35\,200 ${\rm K}$  & 5.61 & -1.50 \\
 	& & & $\pm 500$  & $\pm 0.06$ & $\pm 0.02$ \\
	\hline 
\end{tabular}
\end{center}
\end{table}

We started the analysis by using the metal line blanketed LTE model atmosphere 
grids of O'Toole \& Heber (\cite{otoole}). Using solar metallicity models a 
simultaneous fit of the hydrogen and some helium lines 
(He\,{\sc i} 4026\AA, He\,{\sc ii} 5412\AA\ and in particular He\,{\sc ii} 
4686\AA) was 
not possible (Fit A in Table~\ref{tab:par}). This so called helium problem already 
occurred during other analyses of pulsating subdwarfs 
(Heber et al. \cite{heber2}; Edelmann \cite{edelmann}). The analysis of 
HST-UV spectra of three sdB stars with similar \(T_{\rm eff}\) as 
KPD~1930+2752 revealed supersolar abundances of the iron group 
elements (O'Toole \& Heber \cite{otoole}). Those stars also displayed the 
optical He ionisation problem. Using more appropriate metal-rich models 
(10 times solar metallicity) the problem could be remedied because of a 
modified atmospheric temperature structure due to significantly increased 
line blanketing (see also Heber et al. \cite{heber3}). The abundances of the iron 
group elements could not been determined for KPD~1930+2752 due to its high 
rotational velocity, which causes a strong broadening of all metal lines. 
Because of the similarity of its atmospheric parameters to those of the stars 
studied by O'Toole and Heber (\cite{otoole}), we adopted high-metallicity 
models as well and, indeed, the fit improved 
(Fit B in Table~\ref{tab:par}, Figs. \ref{LTEm0} \& \ref{LTEm1}).

In a second step we used the new grid of hybrid models that account for both 
metal line blanketing and departures from LTE using ATLAS9. We chose 
supersolar metallicity because of the experience with the LTE analysis 
described above. The quality of this fit (labelled C in Table~\ref{tab:par}) is comparable 
to that from the LTE  analysis. 

Then we checked the influence of Balmer line broadening by using synthetic 
spectra calculated from ATLAS9 models, 10 times solar metallicity with the 
hybrid approach and SH broadening tables instead of VCS tables (Fit D in 
Table~\ref{tab:par}). Finally ATLAS9 was replaced by ATLAS12 and the analysis repeated 
(Fit E in Table~\ref{tab:par}) using the same approach as in the previous step.

The resulting parameters from the different grids are summarized in
Table~\ref{tab:par}. 
As can be seen, systematic effects for 
the effective temperature are small, as models A to E differ by $530\,{\rm K}$ 
or less. 
The gravities, however, differ by 
as much as $0.12\, {\rm dex}$. The ``hybrid'' models yield a gravity $0.06\, 
{\rm dex}$ higher 
than the LTE models for supersolar metal content. The replacement of the 
VCS tables by SH tables systematically increases gravity by another 
$0.06\, {\rm dex}$. It is possible that part of the shift is explained by differences
in table organisation and interpolation (Lemke \cite{lemke}).
Note in passing that the solar metallicity LTE model 
(VCS broadening tables), 
i.e. the one of the least sophistication, yields almost identical 
$T_{\rm eff}$ and 
gravity as the most sophisticated model grid 
(supersolar ATLAS12 plus NLTE 
line formation plus SH line broadening). However, the fit of the observed 
spectrum from the latter grid is superior.
It is impossible to judge which of the models in Table~\ref{tab:par} is
 preferable. The quality of fit
lends support for synthetic spectra calculated from supersolar metallicity 
models. We therefore dismissed fit A (calculated for solar metallicity) 
and adopted 
the mean values of the other fits: $T_{\rm eff} = 35\,200 \pm 500\, {\rm K}$ and 
$\log g = 5.61 \pm 0.06\, {\rm dex}$. For sake of completeness we mention that 
the helium abundance differs by $0.11\, {\rm dex}$.
         
It is important to note that the effective temperature is significantly 
higher (about \(2\,000 \, {\rm K}\)) than previously derived in other 
investigations  (Bill\`{e}res et al. \cite{billeres}) irrespective of the 
choice of model atmosphere. Thus, KPD~1930+2752 is situated at the hot edge 
of the
instability strip in the \(T_{\rm eff} - \log \, g\) diagram (Charpinet et al.
\cite{charpinet}).

\section{Projected Rotational Velocity \label{sec:rot}}

The primary aim of the high-resolution time-series spectroscopy was to 
measure 
the projected rotational velocity of KPD~1930+2752 as accurately as possible.
For this purpose, a high S/N spectrum was constructed from the 200 HIRES spectra. 
Beforehand the individual spectra were shifted to rest wavelength. The median 
was calculated in order to filter cosmics. 

The projected rotational velocity 
was measured by convolving a synthetic spectrum calculated from the best fit 
model atmosphere with a rotational broadening ellipse for appropriate 
$v_{\rm rot}\sin i$. The FITSB2 routine of R. Napiwotzki was used, which fits 
the projected rotational velocity and performes an error calculation based on a 
bootstrapping algorithm (Napiwotzki et al. \cite{napiwotzki6}). The lines H$\alpha$, H$\beta$, H$\gamma$, 
He\,{\sc ii} 4686\AA,  He\,{\sc i} 4472\,\AA,  He\,{\sc i} 4922\,\AA, and 
He\,{\sc i} 5016\,\AA\ 
were used simultaneously for this measurement. The 
result is $v_{\rm rot}\sin i = 93.8 \pm 2.2\, {\rm km\,s^{-1}} $. 

\begin{figure}[t!]
	\resizebox{\hsize}{!}{\includegraphics{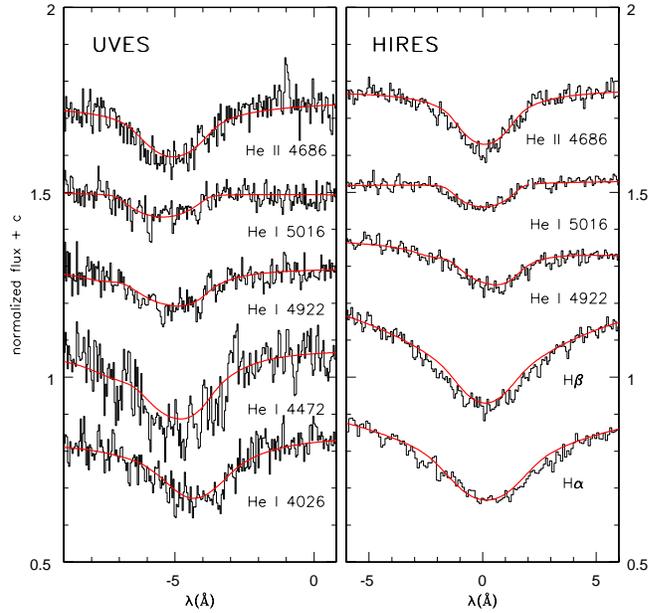}}
	\caption{Best fits of $v_{\rm rot}\sin{i}$ to some 
	lines of the HIRES and UVES spectra. The atmospheric parameters were fixed to 
	the values derived from the CAFOS spectra. LTE models with NLTE line formation 
	and ten times solar metallicity were used (model grid C, Table
	\ref{tab:par}).}
	\label{rot1}
\end{figure}

Fortunately one of the two UVES spectra was taken near the maximum of the 
radial velocity curve (see Fig. \ref{RV2} lower right corner). That means that 
despite the relatively long exposure time orbital smearing is negligible 
and the spectrum can be used for the $v_{\rm rot}\sin{i}$-measurement, too. 
The rotational velocity was obtained as described above from the 
lines H$\alpha$ - H8, He\,{\sc ii} 4686\,\AA, He\,{\sc i} 4026\,\AA, He\,{\sc i} 
4472\,\AA, He\,{\sc i} 4922\,\AA, and He\,{\sc i} 5016\,\AA. 
The measured $v_{\rm rot}\sin i = 90.7 \pm 2.1\, {\rm km\,s^{-1}} $ is 
consistent with the HIRES result. To get a smaller error margin we calculated 
the average of the two independent measurements 
\(v_{\rm rot}\sin{i} = 92.3 \pm 1.5 \, {\rm km\,s^{-1}}\) (see Fig. \ref{rot1}).

\section{Mass and Inclination \label{sec:mi}}

KPD~1930+2752 is obviously affected by the gravitional forces of the companion, demonstrated by its ellipsoidal deformation.
Since the period of the photometric variations is exactly half the period of 
the radial orbit and the two components of the binary are close together the 
rotation of the sdB star is very likely tidally locked to the orbit.
Having determined the gravity and projected rotational velocity, we have 
three equations at hand that constrain the system, with the sdB mass 
\(M_{\rm sdB}\) being the only free parameter.
Besides the mass function

\begin{equation} \label{fm}
	f(M_{\rm sdB},M_{\rm comp})=\frac{M_{\rm comp}^{3}(\sin{i})^{3}}{(M_{\rm sdB}+M_{\rm comp})^{2}}=\frac{K_{\rm sdB}^{2}P}{2\pi G}
	\label{massfunk}
\end{equation}

these are

\begin{equation} \label{sini}
	\sin{i}=\frac{v_{\rm rotsini}P}{2\pi R}
	\label{tlock}
\end{equation}

\begin{equation} \label{rad}
	R=\sqrt{\frac{M_{\rm sdB}G}{g}}
	\label{massrad}
\end{equation}

\begin{figure}[t!]
	\resizebox{\hsize}{!}{\includegraphics{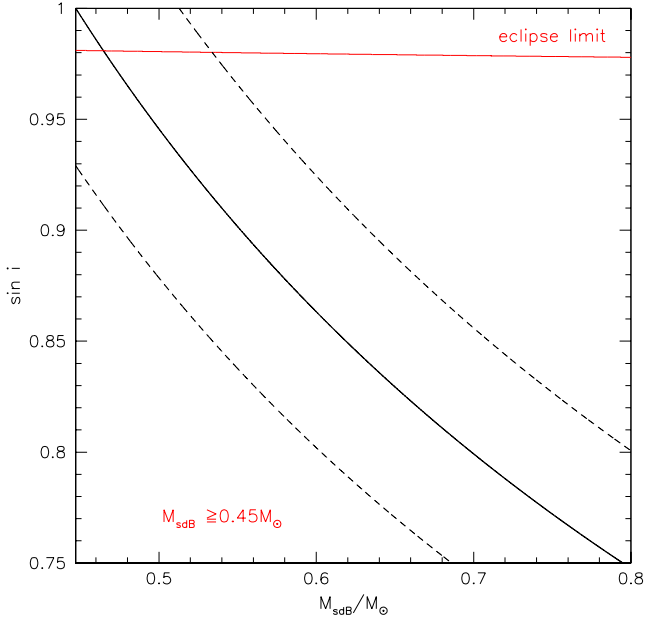}}
	\caption{Inclination versus sdB mass. The dashed lines correspond 
	to $\pm \,1\sigma$ error. The upper curve indicates the minimum 
	inclination for an eclipse. Since $\sin{i}$ cannot exceed unity, a minimum mass for the sdB follows.}
	\label{masssini}
\end{figure}

\begin{figure}[t!]
	\resizebox{\hsize}{!}{\includegraphics{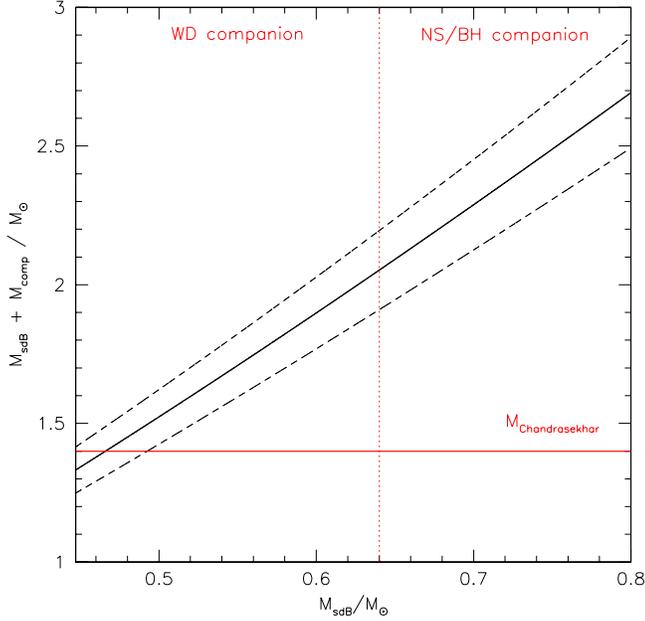}}
	\caption{Total mass of the binary system as a function of the sdB mass with associated error margins (dashed lines) for the total mass. The horizontal line marks the Chandrasekhar limit. The dotted vertical line marks the point where the companion mass equals this limit. For higher sdB masses, the companion cannot be a white dwarf, but has to be a heavier object such as a neutron star (NS) or a black hole (BH).}
	\label{massmass}
\end{figure}

With \(\log \, g\) obtained from the model atmosphere analysis, the radius of 
the star $R$ was calculated using the standard mass-radius relation 
(Eq. \ref{massrad}). 
Together with the orbital period of the system \(P\) and the projected 
rotational velocity $v_{\rm rotsini} = v_{\rm rot}\sin{i}$ the inclination of 
the system $\sin{i}$ was derived for different values of the sdB mass. 
Because the rotation is tidally locked to the orbit, the rotational period of 
the sdB equals the orbital period of the system. Therefore the absolute value 
of the rotational velocity could be calculated. From 
the measured projected rotational velocity the inclination of the system 
was then derived (Eq. \ref{tlock}). With the sdB mass as free parameter, the 
measured radial velocity semiamplitude $K_{\rm sdB}$ and orbital 
period $P$ the mass function was solved 
numerically (Eq. \ref{massfunk}) to derive the mass of the companion 
$M_{\rm comp}$ and calculate the total mass of the binary. The fact that 
\(\sin{i}\) cannot exceed unity gave a lower limit for 
the mass of the sdB $M_{\rm sdB}\geq0.45\,M_{\odot}$ (see Fig. \ref{masssini}). 
The errors were calculated with Gaussian error propagation. 
The error budget is dominated by the error in $\log g$, which has to be 
estimated from the model atmosphere fit. Period $P$, radial velocity 
semiamplitude $K_{\rm sdB}$ and projected rotational velocity 
$v_{\rm rot}\sin{i}$ are measured to a precision more than sufficient for our 
purpose. This means that the accuracy of the results cannot be improved before 
 better model grids become available. The higher value of $\log g$ which 
was obtained by using new line broadening tables indicates a shift of the mass 
minimum and therefore of the total binary mass towards higher values. 
As can be seen in Fig. \ref{massmass} the total mass of the system exceeds the 
Chandrasekhar limit for almost all assumptions of \(M_{\rm sdB}\). 
If the companion is a white dwarf, its single mass has to be lower than the 
Chandrasekhar limit. This implies an upper 
limit \(M_{\rm sdB} \leq 0.64 \, M_{\odot} \) and a possible total mass range 
of \(M_{\rm sdB+WD} = 1.33 \pm 0.08 - 2.04 \pm 0.14 \, M_{\odot} \) (Fig. \ref{massmass}). 
A more massive invisible companion such as a neutron star or a black hole 
cannot yet be 
ruled out completely. In this case the subdwarf mass would exceed the 
possible mass range for this kind of stars (Han et al. \cite{han1}, 
\cite{han2}).
The inclination angle of the system 
(Fig. \ref{masssini}) is close to \(90^{\circ} \). 
KPD~1930+2752 could be an eclipsing binary 
(cf. the very similar binary KPD~0422+5421, Orosz \& Wade (\cite{orosz})). 
The detection of eclipses would rule out any object smaller than a 
white dwarf. If eclipses were detected the sdB mass would be constrained 
to $0.45\,M_{\odot}$ to $0.54\,M_{\odot}$ (see Fig. \ref{masssini}). On the other 
had, if eclipses could be ruled out from high quality light curves, the total 
mass of the system must be larger than $1.55\,M_{\odot}$, well above the
Chandrasekhar limit.

\begin{figure}[t!]
	\resizebox{\hsize}{!}{\includegraphics{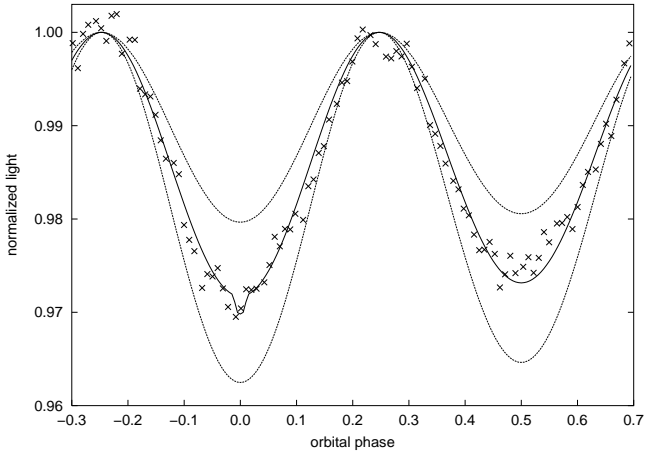}}
	\caption{Light curve data taken from Bill\`{e}res et al. (\cite{billeres}) shown together with examples of synthetic light curves. A good fit ($M_{\rm sdB}\approx 0.47\,M_{\odot}$; solid line) is compared to light curves for $M_{\rm sdB}\approx 0.45\,M_{\odot}$ and $M_{\rm sdB}\approx 0.55\,M_{\odot}$.
	For these two cases the other parameters have been calculated according to Eqns. \ref{fm} to \ref{rad}.}
	\label{fig:lcs}
\end{figure}

\section{Constraints by photometry \label{sec:phot}}

The light curve obtained by Bill\`{e}res et al. (\cite{billeres}) shows 
ellipsoidal variations. We used this information to further constrain the parameters of the
KPD~1930+2752 system. We employed the light curve synthesis and
solution code MORO which is based on the model by Wilson \& Devinney
(\cite{wilson}). The details of the Bamberg implementation are given by Drechsel et al.
(\cite{drechsel}). The software uses a modified Roche model for light curve
synthesis. It is capable to realistically simulate the distortions of the stars induced
by the presence of a companion.

For comparison of the observations with the synthetic light curves of MORO, it
was necessary to prepare the data accordingly. Since the program does not include the
effect of Doppler boosting arising from the high orbital velocity, this effect had to be corrected for. It increases the total flux by a factor of $(1-v(t)/c)^{3}$, where $v$ is the radial velocity at time $t$ and $c$ the speed of light. The boosting is counteracted by the Doppler shift, which reduces the effect by a factor of $(1-v(t)/c)^{2}$ (see Maxted et al. \cite{maxted}). A resulting factor of $(1-v(t)/c)$ was applied to the total flux. Furthermore, the light
curve shows a dip during the maximum at orbital phase 0.25. This effect
is currently unexplained and cannot be modelled by MORO, therefore the
corresponding data points have been left out for the light curve analysis. Because 
KPD~1930+2752 is a pulsating subdwarf with a very rich spectrum of modes, the short period
pulsations have been filtered out by Bill\`{e}res et al. (\cite{billeres}). As this
procedure is not trivial at all, hidden pulsations may limit the accuracy of the results.
One the other hand, the amplitude of the ellipsoidal variations is very sensitive to the 
system parameters as can be seen in Fig. \ref{fig:lcs}. 

\begin{figure}[t!]
	\resizebox{\hsize}{!}{\includegraphics{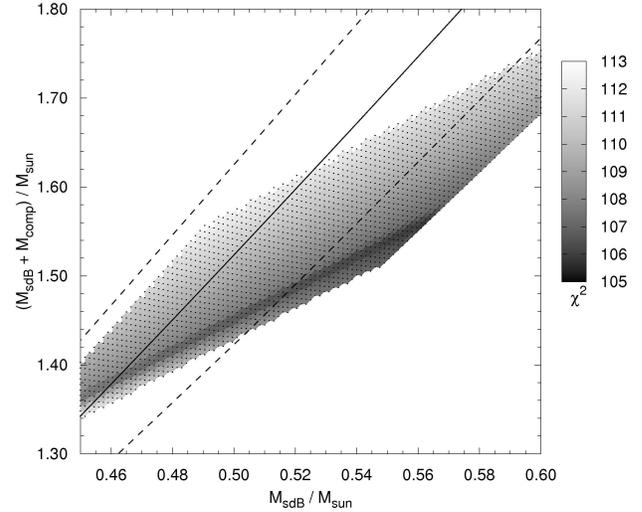}}
	\caption{Detail of Fig. \ref{massmass} with light curve data
	superimposed. The full drawn line is the relation derived from spectroscopy. The dashed lines mark the errors in total mass. Each dot denotes the location of a parameter set for which
	a light curve has been calculated. The background shading corresponds to
	the quality of the fit -- darker shading implying better $\chi^2$.}
	\label{fig:mass}
\end{figure}

At first trial, light curves with different component masses and orbital
inclinations were synthesized. The corresponding changes in the light curve
shapes were significant enough to encourage further investigation. A coarse grid
of light curves was then synthesized which covered the parameter space between
$M_{\rm sdB}=0.45\,M_{\odot}$ and $M_{\rm sdB}=1.0\,M_{\odot}$. For $\sin{i}$ and
$R_{\rm sdB}$ a range of values was calculated for each $M_{\rm sdB}$
according to Eqns. \ref{fm} to \ref{rad}. These ranges covered the parameter
space between the respective error limits derived from spectroscopy. Other relevant parameters
for light curve synthesis like orbital separation $a$ and the companion mass were
derived from Kepler's laws of orbital motion. The radius of the companion was
calculated from the mass-radius-relation for white dwarfs, or set to an
infinitesimal value for neutron star masses (in that case ruling out eclipse effects
in the light curves).
A $\chi^2$ value was calculated for each grid point with respect to the
obsered light curve. A coarse analysis revealed a marked tendency of
good fits to cluster in the $\sin{i}$ regions of marginal eclipse. 
According to Eqns. \ref{sini} and \ref{rad} this behavior strongly favored 
sdB masses close to the canonical value (Fig. \ref{fig:mass}). Lower values of $\sin{i}$ and higher sdB
masses could be ruled out quickly, since they led to quick deterioration of fit quality.

\begin{figure}[t!]
	\resizebox{\hsize}{!}{\includegraphics{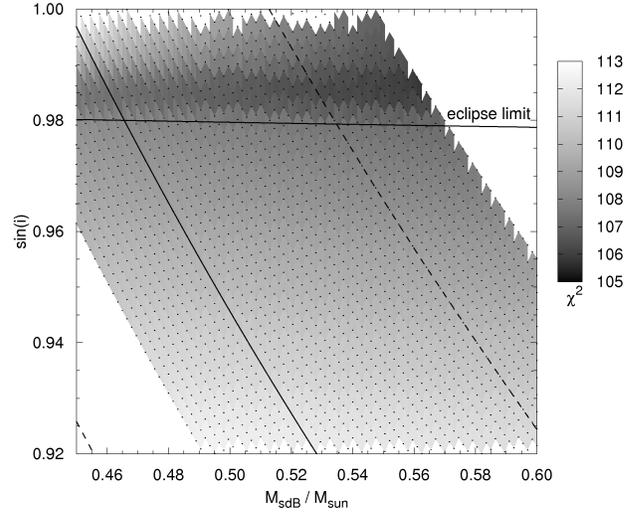}}
	\caption{Detail of Fig. \ref{masssini} with light curve data
	superimposed. The full drawn line is the relation derived from spectroscopy. The dashed lines mark the errors in $\sin{i}$. Each dot denotes the location of a parameter set for which
	a light curve has been calculated. The background shading corresponds to
	the quality of the fit -- darker shading implying better $\chi^2$. Note
	that good solutions cluster slightly above the eclipse limit, indicating a high
	orbital inclination near $80^\circ$.}
	\label{fig:incl}
\end{figure}

In order to probe the relevant portion of the parameter space in greater detail,
a very fine grid of synthetic light curves was constructed,
containing more than 76000 parameter combinations. It
covered only that area of parameter space where promising solutions had been
identified previously. The range of sdB masses extended from $0.45\,M_{\odot}$ to 
$0.60\,M_{\odot}$. Indeed, a significant overlap between the spectroscopically
determined parameter range and a set of very good light curve fits could be
found (see Fig. \ref{fig:incl}). Most notable is the coincidence of the best fits with
the region of eclipse ($i\approx 80\degr$). Fit quality deteriorates significantly for lower inclinations.
Probably this behavior stems from the light curve minimum at orbital phase 0, which shows signs
of an eclipse effect and is therefore fit marginally better by synthetic light curves
displaying the same effect. On the other hand, this marginal dip could also be a remnant from the
removal of the intrinsic stellar pulsations. A better light curve is needed to decide whether the binary is
eclipsing or not.

As noted previously, the light curve fits, favoring an eclipse and therefore a
high inclination, constrain the possible system mass considerably. However the
only absolute parameter which can be determined from the light curves alone is the
orbital inclination, as the results of light curve analyses are well known to be degenerate with respect to the system scale. 
Indeed, the sdB mass is poorly constrained by the light curve fit (see Fig. \ref{fig:incl}) as are the orbital separation and the radii of the stars. However, if we combine the results of the photometric analysis with the mass-inclination relation derived from spectroscopy (Fig. \ref{masssini}), the sdB mass is constrained to a very narrow range of
$M_{\rm sdB} = 0.45 - 0.52\,M_{\odot}$, corresponding to a total mass of $1.36 - 1.48\,M_{\odot}$ (see Fig. \ref{fig:mass}). On the other hand, sdB masses higher
than $M_{\rm sdB}\approx 0.52\,M_{\odot}$ can be discarded since in that part of the parameter
space the spectroscopic mass relation
does not have an overlap with good light curve fits within the error margins. A neutron star
companion can therefore be ruled out.

\section{Conclusion \label{sec:con}}

\begin{figure}[t!]
	\resizebox{\hsize}{!}{\includegraphics{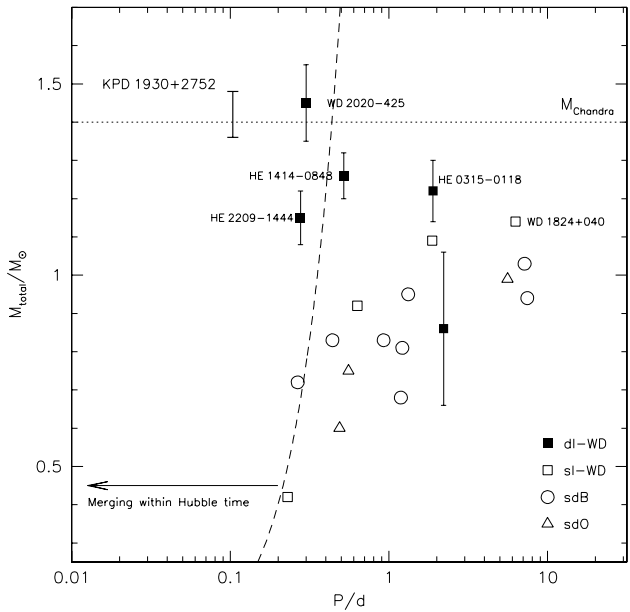}}
	\caption{Total mass plotted against logarithmic period of double degenerate systems from the SPY survey. The filled rectangles mark double lined WDs, for which the absolute mass can be derived. The open symbols mark single lined WDs, sdBs and sdOs. Only lower mass limits can be derived for these binaries by adopting $\sin{i}=1$ (Napiwotzki et al. \cite{napiwotzki3}; Karl \cite{karl}; Napiwotzki et al. \cite{napiwotzki4}; Karl et al. \cite{karl2})}
	\label{Progen}
\end{figure}

The combination of an extensive set of 350 high- and low-resolution spectra,
2\,700 radial velocity data points from prior observations and a reanalysis of the
published light curve made it 
possible to determine the parameters of KPD~1930+2752 with unprecedented 
accuracy. The companion is a heavy white dwarf. 
KPD~1930+2752 is the first subdwarf whose mass could be constrained in this 
way. With $0.45 - 0.52\,M_{\odot}$ it is remarkebly close to the canonical mass of $0.5\,M_{\odot}$.
It is worthwhile to note that masses of other sdBV derived by asteroseismology are also close to the canonical value (Brassard et al. \cite{brassard}; Charpinet et al. \cite{charpinet2}a,b). Furthermore Orosz \& Wade (\cite{orosz}) determine a sdB mass of 
$0.51\pm0.05\,M_{\odot}$ for the eclipsing sdB+WD binary KPD~0422+5421.\\
The total mass and the merging time of the binary indicate that it is 
an excellent double degenerate candidate for an SN~Ia progenitor. 
KPD~1930+2752 is one of only two double degenerate systems, 
which fulfill these requirements 
(see Fig. \ref{Progen}; Napiwotzki et al. \cite{napiwotzki3}). 
While it fits into the DD scenario, KPD~1930+2752 might also evolve into an 
SN~Ia as a single degenerate binary system. 
Since the time for merging due to gravitational 
wave radiation $t_{\rm merger} \approx 2 \cdot 10^{8}\,{\rm yr}$ is of the same order 
as the EHB life time $t_{\rm EHB} \approx 2 \cdot 10^{8}\,{\rm yr}$, Roche lobe overflow 
could occur well before the sdB becomes a white dwarf due to the shrinkage 
of the orbit. Recent calculations by Han and Podsiadlowski (priv. comm.) 
indicate that the mass transfer would be stable. 

Follow-up observations should be undertaken to measure an improved 
light curve and search for clear signs of eclipses. A promising alternative
option is provided for asteroseismology. KPD~1930+2752 is the first pulsating subdwarf with a mass estimate from orbital analysis. It could be used to calibrate asteroseismological models (see e.g. Charpinet et al. \cite{charpinet2}). This is a very difficult task because the frequency spectrum is very complex and
modelling is rendered difficult by the ellipsoidal deformation and the rapid
rotation of KPD~1930+2752.  

\begin{acknowledgements}

We would like to thank C. S. Jeffery, P. F. L. Maxted and J. Orosz for 
providing us with their data. We are also grateful to Z. Han and P. 
Podsiadlowski for their comments and interpretations of our results.
SG was partly supported by the Deutsches Zentrum
f\"ur Luft- und Raumfahrt (DLR) through grant no.\ 50-OR-0202. 
The authors wish to recognize and acknowledge the very significant cultural 
role and reverence that the summit of Mauna Kea has always had within the 
indigenous Hawaiian community.  We are most fortunate to have the opportunity 
to conduct observations from this mountain.

\end{acknowledgements}

\begin{thebibliography}{}

\bibitem[2003]{bennett}
Bennett, C. L., Halpern, M., Hinshaw, G., et al. 2003, ApJS, 148, 1

\bibitem[2000]{billeres}
Bill\`{e}res, M., Fontaine, G., Brassard, P., et al. 2000, ApJ, 530, 441

\bibitem[2001]{brassard}
Brassard, P., Fontaine, G., Bill\`{e}res, M., et al. 2001, ApJ, 563, 1013

\bibitem[1985]{butler}
Butler, K., \& Giddings, J. R. 1985, in Newsletter on Analysis of Astronomical Spectra, No.\,9 (London: Univ. London)

\bibitem[1996]{charpinet}
Charpinet, S., Fontaine, G., Brassard, P., \& Dorman, B. 1996, ApJ, 471, L103

\bibitem[2005]{charpinet3}
Charpinet, S., Fontaine, G., Brassard, P., et al. 2005, A\&A 437, 575

\bibitem[2005]{charpinet2}
Charpinet, S., Fontaine, G., Brassard, P., et al. 2005, A\&A 443, 251

\bibitem[2000]{dekker}
Dekker, H., D'Odorico, S., Kaufer, A., et al., Proc. SPIE, 4008, 534

\bibitem[1986]{downes}
Downes, R. A. 1986, ApJS, 61, 569

\bibitem[1995]{drechsel}
Drechsel, H., Haas, S., Lorenz, R., \& Gayler, S. 1995, A\&A, 294, 723

\bibitem[1990]{dreizler}
Dreizler, S., Heber, U., Werner K., et al. 1990, A\&A, 235, 234

\bibitem[2003]{edelmann}
Edelmann, H. 2003, Ph.\,D thesis, Univ. Erlangen-Nuremberg

\bibitem[1982]{eggleton}
Eggleton, P. P. 1982, ApJ, 268, 368

\bibitem[2001]{ergma}
Ergma, E., Fedorova, A. V., \& Yungelson, L. R. 2001, A\&A, 376, L9

\bibitem[1981]{giddings}
Giddings J. R. 1981, Ph.\,D thesis, Univ. London

\bibitem[1961]{hamada}
Hamada, T., Salpeter, E. E., 1961, ApJ, 134, 638

\bibitem[2002]{han1}
Han, Z., Podsiadlowski, P., Maxted, P. F. L., Marsh, T. R., \& Ivanova, N. 2002, MNRAS, 336, 449

\bibitem[2003]{han2}
Han, Z., Podsiadlowski, P., Maxted, P. F. L., Marsh, T. R., \& Ivanova, N. 2003, MNRAS, 341, 669

\bibitem[1986]{heber1}
Heber, U. 1986, A\&A, 155, 33

\bibitem[2000]{heber2}
Heber, U., Reid, I. N., \& Werner, K. 2000, A\&A, 363, 198

\bibitem[2004]{heber4}
Heber, U., Edelmann, H., 2004, Ap\&SS, 291, 341

\bibitem[2006]{heber3}
Heber, U., Hirsch, H., Str\"oer, A., O'Toole, S., Haas, S., et al. 2006, in Proc. of the Second Meeting on Hot Subdwarf Stars, Baltic Astronomy, 15, 91  

\bibitem[1984]{iben}
Iben, I. Jr., \& Tutukov, A. V. 1984, ApJS, 54, 335

\bibitem[2003]{karl2}
Karl, C., Napiwotzki, R., Nelemans, G., et al. 2003, A\&A, 410, 663

\bibitem[2004]{karl}
Karl, C. 2004, Ph.\,D thesis, Univ. Erlangen-Nuremberg

\bibitem[1997]{kilkenny}
Kilkenny, D., Koen, C., O'Donoghue, D., \& Stobie, R. S. 1997, MNRAS, 285, 640

\bibitem[1997]{koen}
Koen, C., Kilkenny, D., O'Donoghue, D., Van Wyk, F., \& Stobie, R. S. 1997, MNRAS, 285, 645

\bibitem[1993]{kurucz1}
Kurucz, R. L. 1993, Kurucz CD-ROM No. 13 (Cambridge, Mass.: Smithsonian Astrophysical Observatory)

\bibitem[1996]{kurucz2}
Kurucz, R. L. 1996, in Model Atmospheres and Spectrum Synthesis, ed. Adelman, S. J., Kupka, F. \& Weiss, W. W. (San Francisco: ASP), 160

\bibitem[2001]{leibundgut}
Leibundgut, B. 2001, ARA\&A, 39, 67

\bibitem[1997]{lemke}
Lemke, M. 1997, A\&AS, 122, 285

\bibitem[2000]{livio}
Livio, M. 2000, in Type Ia Supernovae: Theory and Cosmology, Cambridge Univ. Press, ed. Niemeyer, J. C., \& Truran, J. W., 33

\bibitem[2000]{maxted}
Maxted, P. F. L., Marsh, T. R., \& North, R. C. 2000, MNRAS, 317, L41

\bibitem[1999]{napiwotzki1}
Napiwotzki, R. 1999, A\&A, 350, 101

\bibitem[2001]{napiwotzki5}
Napiwotzki, R., Edelmann H., Heber U., Karl, C., Drechsel H., et al. 2001, A\&A 378, L17

\bibitem[2002]{napiwotzki4}
Napiwotzki, R., Koester, D., Nelemans, G., et al. 2002, A\&A, 386, 957

\bibitem[2003]{napiwotzki2}
Napiwotzki, R., Christlieb, N., Drechsel, H., et al. 2003, ESO Msngr, 112, 25

\bibitem[2004]{napiwotzki6}
Napiwotzki, R., Yungelson, L., Nelemans, G. et al. 2004, in Spectroscopically and Spatially Resolving the Components of the Close Binary Stars, Proc. of the Workshop held 20-24 October 2003 in Dubrovnik, Croatia, ASP Conference Series, Vol. 318, ed. Hilditch, R. W., Hensberge, H., Pavlovski, K., 402

\bibitem[2005]{napiwotzki3}
Napiwotzki, R., Karl, C. A., Nelemans, G., et al. 2005, in Proc. of the 14th European Workshop on White Dwarfs, ASP Conference Series, Vol. 334, ed. Koester, D., Moehler, S., 375

\bibitem[1999]{orosz}
Orosz, J. A., \& Wade, R. A. 1999, MNRAS, 310, 773

\bibitem[2006]{otoole}
O'Toole, S., \& Heber, U. 2006, A\&A, 452, 579

\bibitem[1999]{perlmutter}
Perlmutter, S., Aldering, G., Goldhaber, G., Knop, R. A., Nugent, P., et al., 1999, ApJ, 517, 565

\bibitem[2004]{przybilla1}
Przybilla, N., \& Butler, K. 2004, ApJ, 609, 1181

\bibitem[2005]{przybilla2}
Przybilla, N., 2005, A\&A, 443, 293

\bibitem[2006]{przybilla3}
Przybilla, N., Nieva, M. F., Edelmann, H., 2006, in Proc. of the Second Meeting on Hot Subdwarf Stars, Baltic Astronomy, 15, 107

\bibitem[1998]{riess}
Riess, A. G., Fillipenko, A. V., Challis, P., et al. 1998, AJ, 116, 1009

\bibitem[1995]{saffer}
Saffer, R. A., \& Liebert, J. 1995, in Proc. 9th European Workshop on White Dwarfs, ed. Koester D., Werner K., Springer Verlag, 221

\bibitem[1999]{stehle}
Stehl\'e, C., \& Hutcheon, R. 1999, \aaps, 140, 93 (SH)

\bibitem[1997]{stobie}
Stobie, R. S., Kawaler, S. D., Kilkenny, D., O'Donoghue, \& D., Koen, C. 1997, MNRAS, 285, 651

\bibitem[2004]{telting}
Telting, J., \& \O stensen, R. 2004, A\&A, 419, 685

\bibitem[1973]{vidal}
Vidal, C. R., Cooper, J., \& Smith, E. W. 1973, \apjs, 25, 37 (VCS)

\bibitem[1994]{vogt}
Vogt, S. S., et al., 1994, Proc. SPIE, 2198, 362

\bibitem[1973]{whelan}
Whelan, J., \& Iben, I. Jr. 1973, ApJ, 186, 1007

\bibitem[1971]{wilson}
Wilson, R. E., \& Devinney, E. J. 1971, ApJ, 166, 605

\bibitem[2002]{woolf}
Woolf, V. M., Jeffery, C. S., \& Pollacco, D. L. 2002, MNRAS, 332, 34

\end{thebibliography}
\end{document}